*machines*

# Discontinuous PWM strategy with frequency modulation for vibration reduction in asynchronous machines

Antonio Ruiz-González [1], Juan-Ramón Heredia-Larrubia[2], Francisco M. Pérez-Hidalgo [1] and Mario Meco-Gutiérrez[1]

[1] Electrical Engineering Department, University of Malaga EII, Doctor Ortiz Ramos s/n 29071 Málaga, Spain
[2] Electronic Technology Department, University of Malaga EII, Doctor Ortiz Ramos s/n 29071 Málaga, Spain
* Correspondence: Juan-Ramon Heredia-Larrubia (e-mail: jrheredia@uma.es).

**Abstract:** The aim of this paper is the reduction of vibrations in induction motors. For this purpose, a discontinuous pulse width modulation (PWM) control strategy based on carrier wave modulation applied to multilevel inverters is proposed. This work justifies the reduction of vibrations in the machine in comparison with existing control techniques in the technical literature. The proposed technique also has the feature of attenuating the Total Harmonics Distorsion of the voltage of the multilevel inverters, as well as achieving a higher RMS value of the output voltage for the same DC level. The control strategy allows the electrical spectrum to be varied given a number of pulses per period by changing a parameter of the carrier wave, thereby avoiding natural mechanical resonance frequencies. In this way, motor vibrations can be reduced. Laboratory results for an induction motor with different modulation strategies applied in a multilevel inverter and compared to the strategy presented are attached.

## 1. Introduction

The Induction motor is the most widely used mechanical drive equipment in industrial environments and has positioned itself as an important component in fields such as intelligent transport and machine manufacturing [1]. These devices need to be controlled using power converters that regulate their torque and speed. These converters generate a pure non-sinusoidal power supply that produces electrical harmonics. These harmonics produce vibrations and noises in the motors that can deteriorate the motor and reduce its life cycle [1, 2]. The proposal of this work is to present a possible solution to avoid these harmonics and therefore vibrations in the motor, for which the use of multilevel power converters (MLI) together with an optimised control technique is proposed.

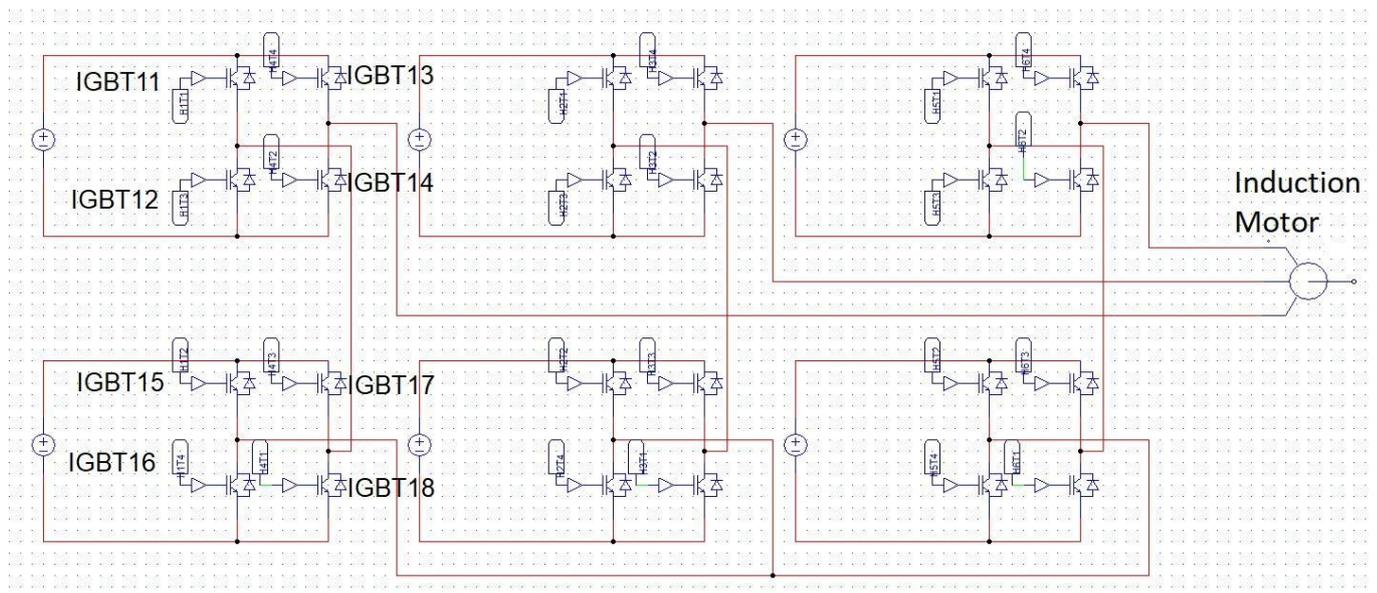

(a)

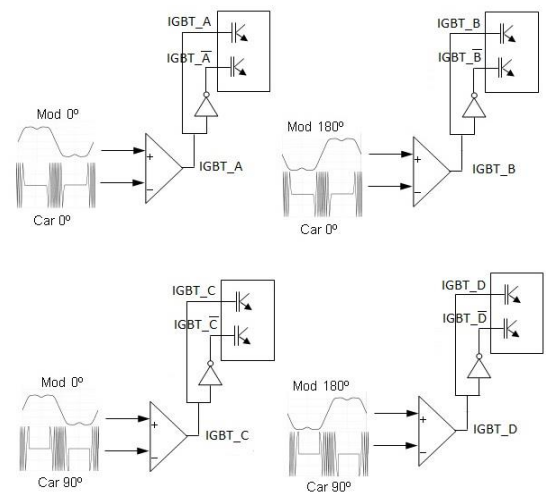

(b)

**Figure 1.** (a) CHB Multilevel power inverter. (b) CHB Control of the signals.

The multilevel inverter is a power converter that provides a suitable solution for synthesizing an output voltage to reduce the harmonic content of voltage and current in medium and high power systems [3]. The most common topologies of these converters consist of connecting individual inverters called stages to provide an output voltage that is the result of associating the voltages of the different stages. The advantages of multilevel converters over two-level converters are lower dv/dt, lower switching losses, higher fault tolerance capability and better output waveform quality [4]. Due to this last advantage the output filter will also have a smaller volume.

There exists a broad range of multilevel inverter topologies [5,6], with the most prevalent voltage source topologies being the neutral point diode (NPC) converter [7,8,9,10], the flying capacitor converter (FCC) [11], the cascaded H-bridge (CHB) converter [12,13], and the modular multilevel converter (MMC) [14,15,16].

In this research paper, we have opted for a cascaded H-bridge multilevel inverter for the comparative analysis (Figure 1.a), which boasts several advantages over alternative topologies (NPC, FCC). The CHB inverter possesses modularity and enhanced controllability due to the identical structure of each stage. Furthermore, it enables the attainment of multiple levels using minimal components. These alleviate switching losses in the utilized devices, thereby augmenting circuit reliability and efficiency [17]. Consequently, this topology finds extensive utilization in industrial applications [6].

One of the applications of the CHB inverter is the provision of power to high-power motors, which are supplied at elevated voltages to minimize current in the windings. This necessitates multiple stages for each phase. It is widely recognized that using such devices for speed or torque control of AC electrical machines can lead to increased vibrations.

Regarding modulation techniques, the literature offers numerous options for MLI control. The most commonly employed techniques include carrier-based PWM [18], space vector modulation [19,20], and selective harmonic elimination [21,22]. Carrier-based modulation gives rise to highly popular techniques for MLIs, such as carrier phase shift pulse width modulation (PS-PWM) and carrier level shift pulse width modulation (LS-PWM) [23,24]. Other techniques involve modifying the carrier signal frequency and modulating signal harmonics' amplitudes to reduce losses and enhance electrical parameters of the modulated waveform [25]. In [26], an LS-PWM technique is utilized, where the reference signal is vertically shifted to derive the switching signal for the MLI. As there exists a level shift between the carrier signals, higher voltage distortion occurs, which is not the case with PS-PWM.

In Figure 1.b), the signals driving the gates of a 2-stage H-bridge are illustrated. The pulse train that triggers the base of IGBT-A is generated by comparing the modulating signal with the 0° phase harmonic injection to the 0° phase carrier. Similarly, the pulse train that activates the base of IGBT-B is obtained by comparing the modulating signal with the 180° phase harmonic injection to the 0° phase carrier. IGBT-C is stimulated by the pulses resulting from the comparison between the modulator and carrier of phases 0° and 90°. Lastly, the 180° phase modulator compared to the 90° phase carrier generates the pulses that drive the gate of IGBT-D.

In summary, all electric drives contribute to increased vibrations in electromagnetic circuits. However, the level of these vibrations can be mitigated depending on the control technique employed [27,28,29]. The proposed strategy focuses on cascaded or H-bridge inverters and utilizes an efficient PS-PWM technique. This technique not only ensures a balanced power distribution among the cells but also results in reduced voltage distortion and effective suppression of harmonic current distortion. These characteristics play a crucial role in attenuating the vibrations generated by the motor. The vibration results and quality parameters of the inverter output waveform obtained through the proposed modulation technique will be compared with results from other techniques documented in the scientific literature, which have shown promising outcomes. Specifically, the compared techniques include amplitude-shifted modulation in the carrier waves (Figure 2a), phase-shifted modulation in the carrier waves (Figure 2b), and phase-shifted modulation with harmonic injection into the modulating wave (Figure 2c). To the best of our knowledge, there is currently no recent research in the technical literature on reducing vibrations in induction motors using multilevel converters and PS-PWM techniques.

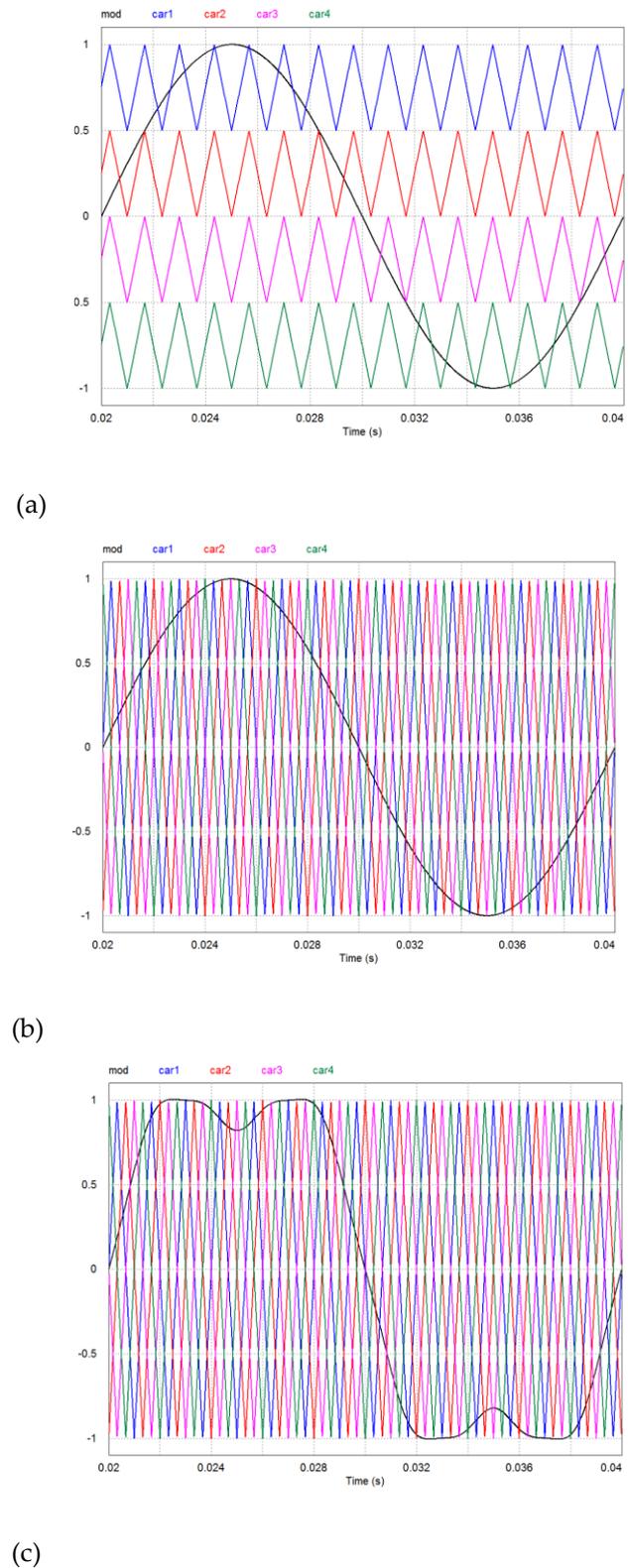

**Figure 2**. Multilevel inverter modulation: a) with amplitude shift; b) with phase shift; c) with phase shift and injection of harmonics.

The paper is structured as follows in the subsequent sections. Section 2 provides an introduction to the anticipated vibration frequencies when feeding an induction motor with either a sinusoidal source or an inverter generating specific sets of harmonics.

In Section 3, the proposed technique for controlling multilevel inverters is presented, outlining its fundamental principles. The generation of the signals applied to the IGBT gate terminals and the waveform at the inverter's output, based on the control parameter, is described. Additionally, the section illustrates the relationship between the instantaneous frequency of the carrier wave and the modulating signal in relation to the control parameter.

Moving on to Section 4, it presents and analyzes the laboratory results. It showcases the equipment required to generate the control waves for the tested strategies, including the vibration meter setup and the signal analyzer used to record voltage and current waveforms along with their corresponding electrical spectra. Lastly, Section 5 summarizes the conclusions drawn from the research.

**2. Natural frequencies and axial stress generating MMFs on the stator**

There are two sources of vibrations in an electrical machine depending on their origin. The first type is composed of axial forces of aerodynamic and mechanical origin, and by the exentricities of its axis (fan, bearing, activation of mechanical resonances and phenomenon of magnetostriction, mainly). They are low frequency, depend on the degree of saturation in the magnetic circuit of the machine and the load level has limited effect on them. Electrical harmonics of the spatial type and due to the restriction of the maximum number of slots and coils (size function) are another source of vibrations and noise. On the other hand, when an inverter is used to power the motor, its electrical harmonics produced are a new source of vibrations. This last source of vibration will be a function of the modulation technique used with the power switches of these inverters, the carrier frequency, and the speed set for the motor shaft. These vibrations depend on the load level. When any of these electrical harmonics activate a natural resonance frequency of the engine, the level of vibration can be critical to the lifetime of the machine.

The calculation of the natural resonance frequencies of an induction motor structure can be carried out with the equations found in the technical literature [30]. If the stator length is greater than its mean diameter, $D_c$, good results are obtained for these frequencies by considering the electrical machine as a cylinder of infinite length, which for circumferential modes of vibration m≥0 the frequency can be expressed as:

$$f_m = \frac{P_m}{\pi D_c} \sqrt{\frac{E_c}{\rho_c(1-\nu_c^2)}} \quad (1)$$

where $\nu_c$ is Poisson's ratio, $E_c$ is modulus of elasticity, and $P_m$ the roots of the equation of movement. Circumferential mode forces m to zero, and therefore, $P_m$=1. For m≥1

$$P_m = \frac{1}{2}\sqrt{(1+m^2+k^2m^4) \pm \sqrt{(1+m^2+k^2m^4)^2 - 4k^2m^6}} \quad (2)$$

and the dimensionless thickness parameter is:

$$\kappa^2 = \frac{h_c^2}{3D_c^2} \quad (3)$$

being $h_c$ the stator thickness. On the other hand, a housing with a bell-shaped end behaves like a cylinder with both ends mechanically constrained. There will be two axes of displacement: a radial one with circumferential vibrational modes m, "breathing" m=0, elliptical, m=1, etc.) and another axial one with modes n equal to 1 onwards. The characteristic equation of the movement of a cylinder of finite length $L_f$ supported on legs according to the Donnell-Mushtari theory, will be of the following type:

$$P_{mn}^6 - C_2 \cdot P_{mn}^4 + C_1 \cdot P_{mn}^2 - C_0 = 0 \tag{4}$$

where $P_{mn}$ are the roots of the equation of motion. The housing resonance frequencies have the same structure as in equation (1). Three groups of roots correspond to the displacement in the 3 orthogonal directions in Equation (4), for which the smallest real root determines the natural deflection frequency of the frame being:

$$\begin{aligned}
C_2 &= 1 + \frac{1}{2}(3-\nu)(m^2 + \lambda^2) + \kappa^2(m^2 + \lambda^2)^2 \\
C_1 &= \frac{1}{2}(1-\nu)[(3+2\nu)\lambda^2 + m^2 + (m^2+\lambda^2)^2] + \frac{3-\nu}{1-\nu}\kappa^2(m^2+\lambda^2)^2 \\
C_0 &= \frac{1}{2}(1-\nu)[(1-\nu^2)\lambda^4 + \kappa^2(m^2+\lambda^2)^4] \\
\lambda &= 0.5n\pi \frac{(D_f - h_f)}{L_f - L_0} \\
\kappa &= \frac{h_f^2}{12R_f^2}
\end{aligned} \tag{5}$$

$$L_0 = L_f \frac{0.3}{n+0.3} \tag{6}$$

where $R_f$ is the average radius of the casing, $D_f$ is the average diameter of the casing, $h_f$ is the thickness of the casing, $L_f$ is its length, $\nu$ is its Poisson's ratio, and $\rho$ is the density. The circumferential vibrational modes: m iqual to 0 onwards (natural numbers) should be calculated for each axial vibrational mode n.

If the motor is fed from a balanced three-phase power supply, the only harmonics that appear are spatial harmonics. The most significant of these are the spatial harmonics are the tooth harmonics. They are present when the machine is fed with a three-phase network and depend on the number of phases and poles and the winding factor of each harmonic:

$$\nu = k\left(\frac{s_1}{p}\right) \pm 1 \tag{7}$$

being p, the number of pair of poles; $s_1$ the stator slots, and k, natural numbers from 1 onwards. The amplitude of the vibrations A shall be proportional to the products of the terms of the flux densities:

$$A = f(B^2(\alpha, t)) \tag{8}$$

where B is the magnetic flux density in the magnetic circuit of the machine; $\alpha$, angular position with respect to a reference and t, time; so that the amplitude and frequency of these vibrations must consider the product of all harmonic terms of the induction in the machine. Other harmonics are due to the phenomenon of magnetostriction and bearing rollers.

When the machine is powered by a power inverter, the radial forces resulting from the electrical harmonics are caused by the interaction between the stator harmonics and their corresponding time harmonics, resulting in a specific frequency:

$$f_r = 2f \cdot (2km_1 \pm 1) \tag{9}$$

where f is the frequency of the fundamental component of the modulated waveform, k is the series of natural numbers, and $m_1$ is the number of phases of the stator. Similarly, the harmonic decomposition of the rotor currents will result in vibrations at specific frequency values:

$$f_r = 2f \cdot (2km_1 \pm 1) \cdot \left(\frac{s_2}{p} \pm 1\right) \tag{10}$$

being $s_2$ the number of rotor slots. In addition, frequency vibrations are generated resulting from the mutual influence between the frequency of the carrier signal, $f_c$, with the time harmonic frequency of the modulating wave. Its vibrational mode is zero and therefore, the possibility of producing high vibrations is greater since these are reduced with the order of the vibrational mode. Their frequencies can be obtained as:

$$f_r = \left| \pm (n \cdot f_c \pm n' \cdot f) - f \right| \tag{11}$$

where $f_r$ is the frecuency of the force density. If n is even, n' is odd and if n is odd, n' will be even. Therefore, electrical harmonics can be expected at the following frequencies: $f_c \pm 2f$, $f_c \pm 4f$, $2f_c \pm f$, $2f_c \pm 3f$ and so on. In other words, it can be anticipated that the vibrational harmonics generated by the inverter will manifest themselves at frequencies that are either slightly above or below the frequency of the electrical spectrum of the modulated wave applied to the electric motor. When an electrical harmonic is present at a frequency of f, the corresponding vibrational harmonic will manifest itself at a frequency of either (f+1) or (f-1). This is due to the axial force densities being derived from the multiplication of all the flux densities within the air gap.

Power inverters manipulate the voltage level and frequency in the stator of the machine, resulting in an electrical spectrum characterized by amplitudes and frequencies that are contingent upon the modulation technique employed. The existence of these harmonics will produce energy losses, reduced efficiency, and pulsating torques that translate into vibrations in the machine. One common approach to mitigate the adverse effects of these electrical harmonics is to raise the frequency of the carrier signal. However, this frequency is constrained by the maximum switching frequency of the circuit breakers and, in turn, increases the losses incurred by the inverter. Another approach, enabled by the decreased cost of power switches, is to interconnect multiple elementary stages in series, each utilizing one or more separate sources. This arrangement reduces the number of switchings for each switch and enables a higher carrier frequency in the PWM technique, thereby amplifying the voltage and power capacities of the electrical machine.

In the case of high power and voltage requirements, minimizing switching losses becomes crucial. This is achieved by minimizing the frequency of the carrier wave in the inverter modulation. However, this strategy amplifies the likelihood of mechanical resonances in the system caused by the time harmonics generated by the inverter. Hence, for such applications, the control strategy assumes special significance.

## 3. New HIPWM-FMTC strategy applied to multilevel inverters.

The HIPWM-FMTC (Harmonics Injection PWM - Frequency Modulated Triangular Carrier) technique generates the output voltage of an inverter by comparing a modulating wave that injects harmonics with a triangular carrier wave that is frequency-modulated.

The frequency of the carrier waveform is not constant but varies based on a periodic function, specifically a squared cosine, which is synchronized with the phase of the modulating waveform. The strategy employed involves increasing the number of switching pulses when the slope of the modulating waveform is at its maximum or minimum, and gradually reducing it to zero when the slope is at its lowest (near the maximum and minimum phases). However, the total number of pulses in a complete period is maintained in relation to any conventional PWM technique. This strategy is commonly referred to in technical literature (particularly in the context of 3-level inverters) as HIPWM-FMTC (Harmonic Injection PWM - Frequency Modulated Triangular Carrier) [31,32,33,34,35,36].

A modified version of the HIPWM-FMTC technique has been proposed, which introduces a parameter that interrupts the triangular carrier wave during certain time slots of the modulating wave. This modification allows for an increase in the instantaneous switching frequency in some time slots while eliminating switching in others [33]. When applied to multilevel inverters, this technique is referred to as HIPWM-FMTC truncated or HIPWM-FMTCt [35]. The instantaneous pulsation of the carrier wave, denoted as $\omega_i$, becomes a discontinuous function that is synchronized with the modulating wave $\omega_m$. It can be defined as follows:

$$\omega_i = \frac{d\theta}{dt} = A_M \cdot \omega_m [cos^2 \omega_m t - K] \tag{12}$$

The modulating wave and the instantaneous pulse of the carrier wave, $\omega_i$, need to be synchronized functions, as depicted in Figure 3. The parameter K, which represents the truncation level, is a real number ranging from 0 to 1, and it is independent of $\bar{M}$, which represents the number of pulses per period. On the other hand, $A_M$ is a parameter that determines the middle of the maximum frequency of the carrier wave. By selecting appropriate values for $A_M$, K, and the frequency modulation order, $\bar{M}$, the instantaneous maximum frequency can be adjusted. The combination of these parameters ($A_M$ and K) enables the modification of the electrical spectrum without altering the value of $\bar{M}$. Thus, after setting a value for $\bar{M}$, each value of K will correspond to a specific value of $A_M$. It is important to note that when Equation (12) yields negative values, $\omega_i$ will be nullified during those intervals.

Note that the frequency modulation order, $\bar{M}$ is the mean value of a periodic function M(t), thus, the instantaneous frequency modulation order, and which will be null between $t_1$ and $t_2$, and between $t_3$ and $t_4$ for each period of the modulating wave to avoid such switching. During a period of the modulating wave, both M and $\bar{M}$ will be the same value:

$$\bar{M} = \frac{1}{T_m} \int_0^{T_m} A_M [cos^2(\omega_m t) - K] dt \tag{13}$$

where $T_m$ is $2\pi/\omega_m$. The frequency modulation order $\bar{M}$ must be set to a natural number, and if the motor is 3-phase, it must be set to an odd value multiple of 3.

The primary objective is to ensure that the maximum instantaneous frequency of the triangular carrier wave aligns with the points where the modulating sine wave exhibits the steepest slope (0 and π radians). The frequency of the carrier wave gradually decreases following a quadratic cosine function around these phases and eventually reaches zero within a specific time interval determined by the truncation factor, as described by Equations (12) and (13). In the case of K = 0, the carrier wave frequency theoretically becomes zero only at the phase values of the modulating wave: π/2 and 3π/2. In other words, the instantaneous pulsation of the carrier wave is never eliminated. As K increases from 0 to 1 (excluding 1, which would imply an infinite instantaneous carrier wave frequency), the proposed technique ensures that the carrier frequency becomes zero for progressively longer time intervals centered around the T/4 and 3T/4 of the period of the modulating wave. The frequency of the carrier waveform increases precisely around the 0 and T/2 of the period of the modulating wave, which correspond to the points of its steepest slopes. This approach is adopted because the modulating wave undergoes faster slope changes at these phase values, thereby providing a more accurate representation of the carrier wave's shape in the modulated waveform. As shown in Figure 3, K = 0.55 and $\bar{M}$ =15 ($A_M$ = 111.15125), for slightly less than half of the period there will be modulated triangular carrier, and during the other slightly more than half of the period, switching will be canceled. The maximum excursion of the carrier frequency will be $A_M \cdot \omega_m \cdot (1-K)$ = 50.01806·$\omega_m$. Therefore, $t_1$ is 2.3426 ms; and $t_2$, $t_3$ y $t_4$ will be 7.6574, 12.3426, and 17.6574 ms, respectively. For other values of K, regardless of the value of M, the times $t_1$ to $t_4$ will change. For example, for K=0.5, they will be 2.5, 7.5, 12.5, and 17.5 ms.

By employing this technique, the maximum frequencies of the carrier will be π times $\bar{M}$, instead of $\bar{M}$ as would be the case in a Sinusoidal PWM technique. In Figure 3, the area beneath the curve of the instantaneous modulation order corresponds to $\bar{M}$ = 15 and K=0.55. It is possible to identify the modulating, carrier, modulated functions, as well as the function of the instantaneous modulation order. In Figure 4, the combined modulating and carrier functions can be observed as a function of time.

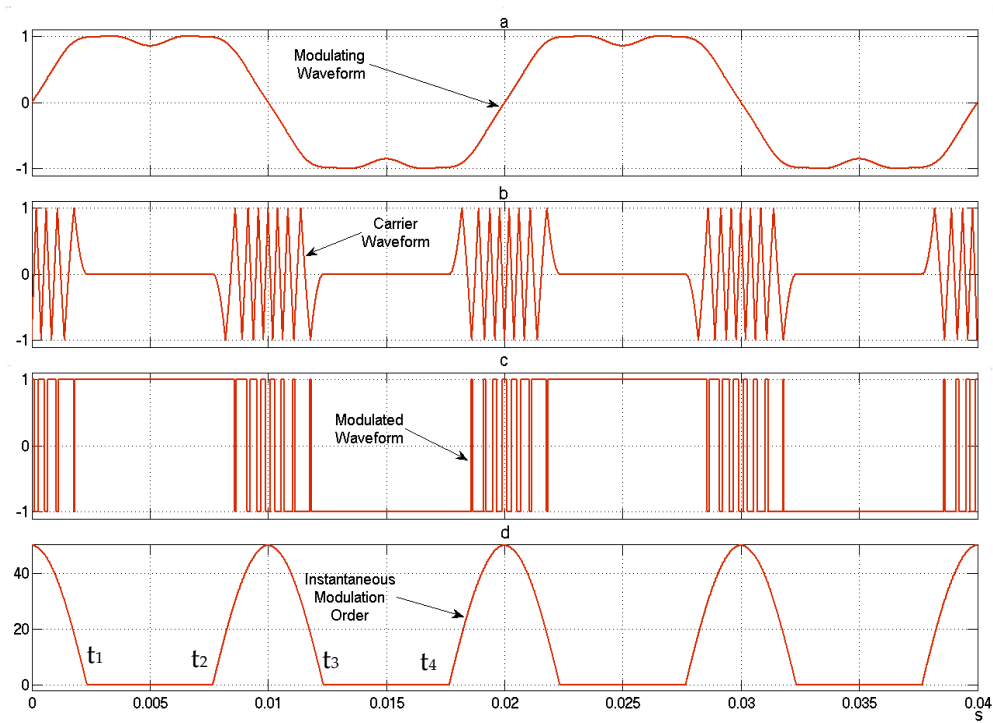

**Figure 3.** Simulation of the proposed PWM generation. a) Modulating waveform for f = 50 Hz; b) HIPWM-FMTCt carrier (K = 0.55 and $A_M$ = 111.15125); c) Phase voltage for an H-bridge; Instantaneous modulation order.

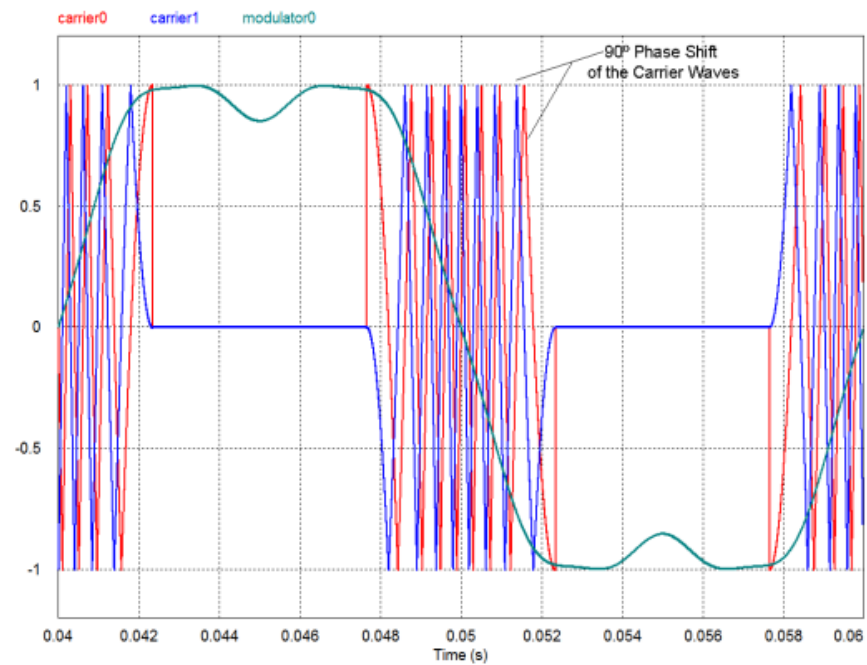

**Figure 4**. Modulator and carrier waveforms for f = 50 Hz, $\overline{M}$=11 for the proposed technique (K = 0.55 and $A_M$ = 22π).

Table I shows how some variables evolve against K when $\omega_m$ are set to 100π rad/s while $\overline{M}$ is fixed at 11 and 15, respectively. The variables presented are $t_1$, (a quarter of the time that makes possible per-period switching of the output waveform); $A_M$; and the instantaneous modulation order of maximum frequency, $A_M \cdot (1-K)$. The maximum carrier frequency is $f_m$ times $A_M \cdot (1-K)$. Figure 5 shows the simulation of two periods of the three-line voltage of the modulated waves, where it can be seen how the switching frequency is higher for values close to zero, and it decreases until the commutations in the maximum and minimum values disappear.

Table 1. $A_M$, $t_1$ and $A_M(1-K)$ for different values of K with M=11 and M=15

| K | $t_1$ (ms) | $A_M$ M=11 | $A_M \cdot (1-K)$ M=11 | $A_M$ (M=15) | $A_M \cdot (1-K)$ M=15 |
|---|---|---|---|---|---|
| 0.2 | 3.5242 | 32.4700 | 35.4218 | 44.27732 | 35.4218 |
| 0.3 | 3.1550 | 40.4314 | 38.5935 | 55.13370 | 38.5935 |
| 0.4 | 2.8207 | 51.8016 | 42.3831 | 70.63850 | 42.3831 |
| 0.45 | 2.6599 | 59.4751 | 44.6063 | 81.10240 | 44.6063 |
| 0.5 | 2.5000 | 22π | 47.1239 | 30π | 47.1239 |
| 0.55 | 2.3426 | 81.5109 | 50.0181 | 111.1513 | 50.0181 |
| 0.6 | 2.1835 | 97.9098 | 53.4054 | 133.5134 | 53.4054 |
| 0.7 | 1.8480 | 152.6484 | 62.4471 | 208.1569 | 62.4471 |
| 0.8 | 1.5153 | 283.9854 | 77.4506 | 387.2528 | 77.4506 |

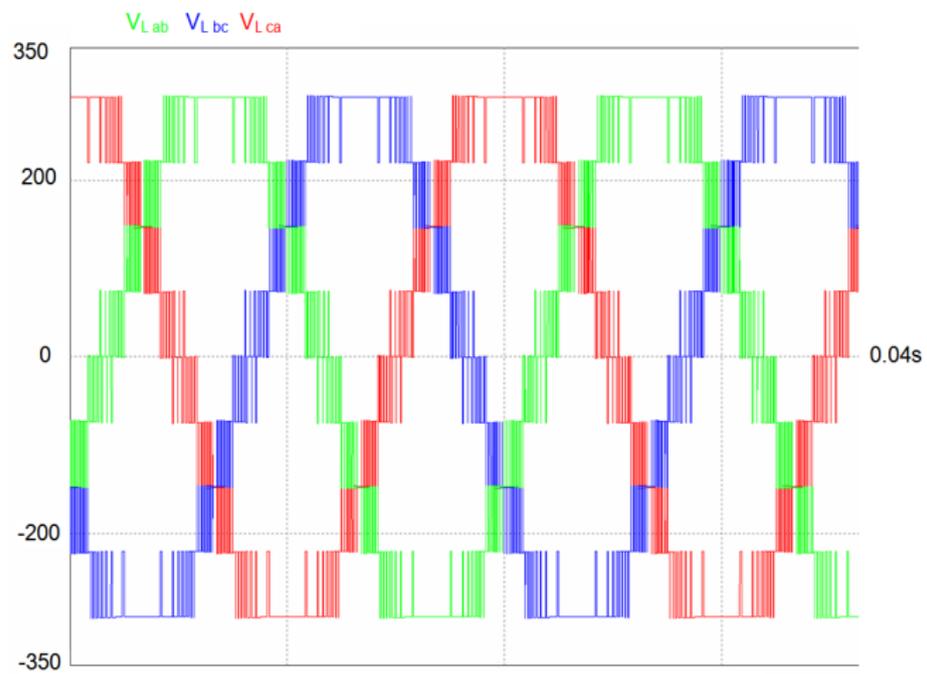

**Figure 5.** Modulated line voltage waveforms with HIPWM-FMTCt strategy (K=0.55, f=50Hz and $\bar{M}$ =15).

## 4. Results

An AEG™ 380/220 V, 1000 W, and 4 poles asynchronous motor with $s_1$ = 36 slots in the stator and $s_2$ = 26 slots in the rotor was used to measure vibrations in the laboratory. The multilevel inverter was built on the basis of GUASCH S.A™ modules, which have GPT-IGBTs transistors. The GUASCH S.A™ GPT-IGBT module was used to implement the multilevel inverter. It offers a power stack of insulated gate bipolar transistors (IGBTs) for motor control. The power system consists of three-phase bridge rectifiers, capacitor banks, IGBTs with forced air-cooled heat sink, opto-coupled drivers, output phase current sensors, DC-Link current and voltage sensor. The maximum voltage that the DC-Link can withstand must not exceed 750 V, and the maximum RMS current in each phase is 32 A.

The hardware system for generating control signals for each of the H inverters that make up the multilevel inverter is based on the NI9154 card from National Instruments™, on which a Labview™ platform has been developed for generating the PWM techniques that control the different H bridges of the multilevel inverter. Chauvin Arnoux™ C.A 8336 was used to measure the output current harmonics of the multilevel inverter. To measure the vibrations of the machine, a sensor PCB™ Piezotronics Accelerometer model Number 333B50 has been connected to a METRAVIT Symphonie model 01 dB sound level meter. Figure 6a illustrates the multilevel power inverter, the NI9155 control equipment and Figure 6b illustrates the sensor on the AEG™ asynchronous motor.

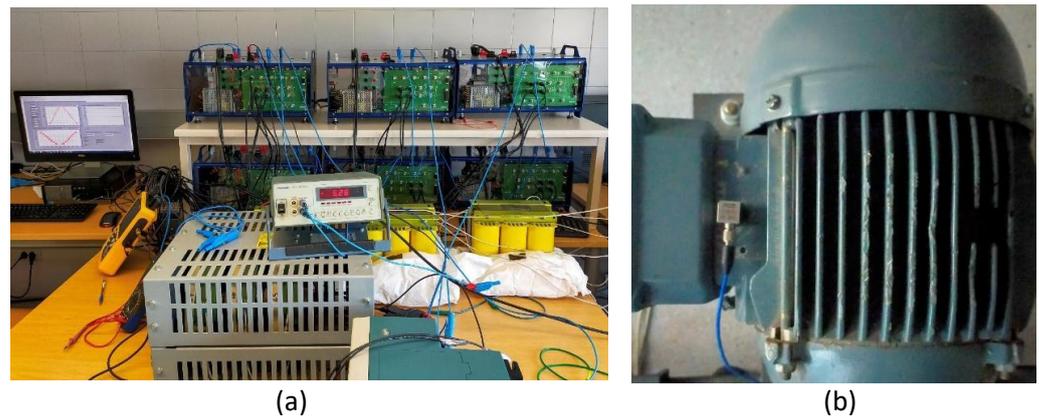

(a) (b)

**Figure 6.** Implementation: a) Multilevel inverter, NI9154 network analyser, b) Motor with PCBTM piezoelectric accelerometer model 333B550

The Figures 7, 8, 9, and 10 are the waveforms and amplitude of current spectra (%) of the different PWM techniques: SPWM-I, SPWM-II, SPWM-III techniques and the HIPWM-FMTCt technique with K=0.55 presented in this paper.

The SPWM-I technique corresponds to PWM with amplitude-shifted triangular carrier wave. This is perhaps the most common technique in the technical literature on multilevel inverters. Figure 7a shows the output waveform of the multilevel inverter and Figure 7b the harmonics electrical spectrum of the output current of this waveform. The fundamental harmonic in percentage terms of the output voltage is maintained at 220 V RMS.

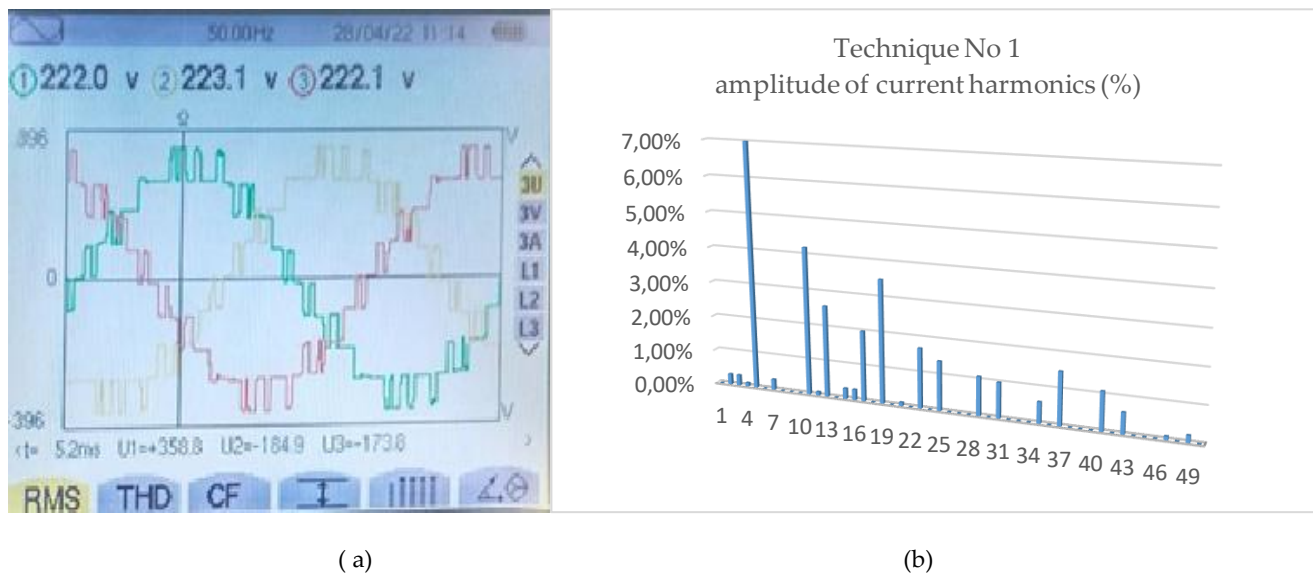

( a) (b)

**Figure 7.** a) Line voltage output with SPWM-I strategy; b) Amplitude of current harmonics (%) V=220 V RMS of the fundamental term.

The following SPWM-II technique corresponds to phased-shift for the triangular wave carriers, and sine modulator wave. As in the previous technique in Figure 6a and Figure 6b corresponds to the inverter output voltage waveform and the harmonics electrical spectrum in percentage terms of the current respectively. The output voltage is maintained at 220 V RMS.

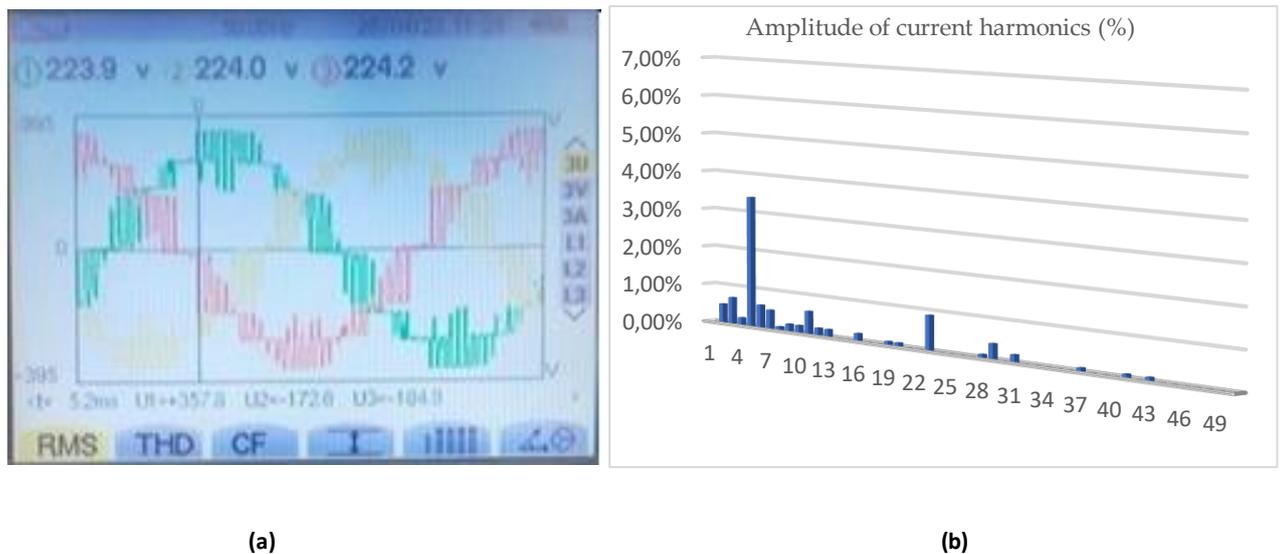

(a)  (b)

**Figure 8.** a) Line voltage of the multilevel inverter with SPWM-II strategy; b) Amplitude of current harmonics (%): V=220 V RMS of the fundamental term.

The SPWM-III technique uses a triangular carrier and harmonics are injected into the modulating waveform. Figure 9a shows the modulated waveforms for the line signal of the multilevel inverter with SPWM-III and Figure 9b shows the electrical spectrum of the output current (%) of the SPWM-III multilevel inverter. (V=220 V RMS of the fundamental term).

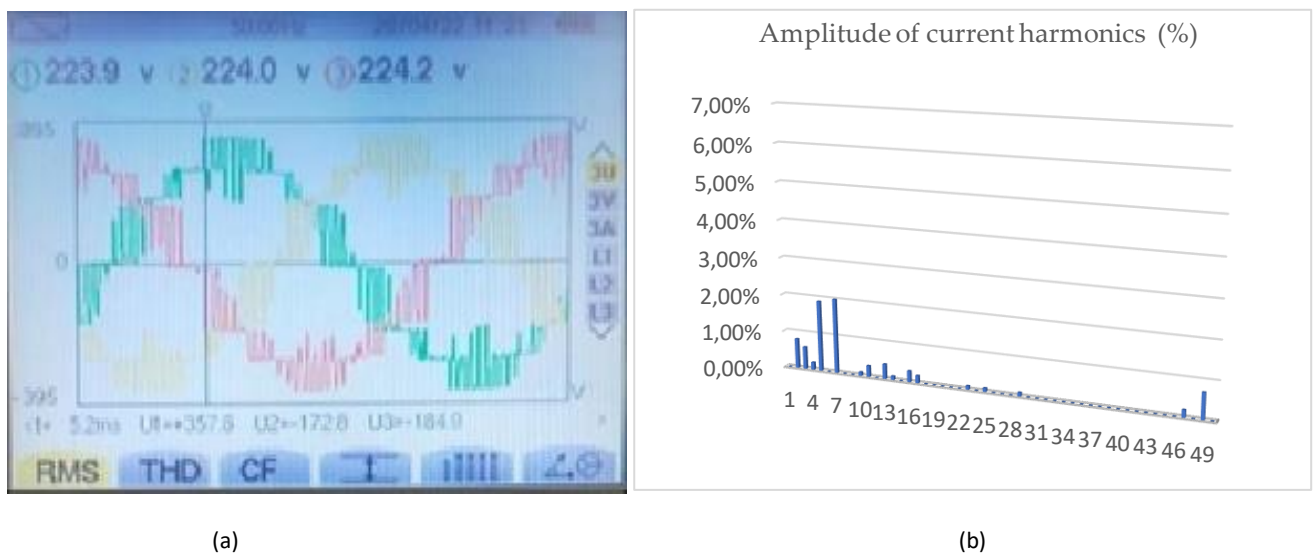

(a)  (b)

**Figure 9.** a) Line voltage output of the multilevel inverter with SPWM-III technique; b) Percentage amplitude of current harmonics. V=220 V RMS of the fundamental term.

The proposed technique is HIPWM-FMTCt which has the characteristic of modulating and truncating the frequency of the carrier wave. Figure 10, in sections a and b, shows the output form the inverter voltage and the percentage electrical spectrum of the output current as in the previous techniques. The same value of the base term is maintained as in the previous techniques at 220 V RMS. For these graphs a K=0, 55 has been chosen, as this is the best value of k to reduce vibrations within the field of measurements made.

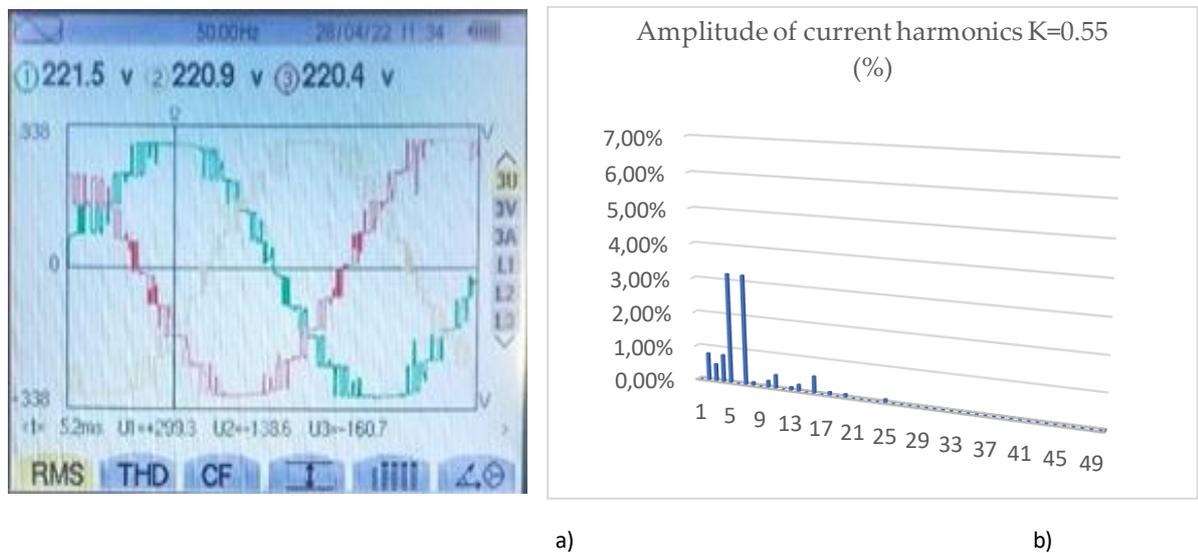

a)  b)

**Figure 10.** a) Line voltage output of the multilevel inverter with technique HIPWM-FMTCt with K=0.55 and M=15; b) Percentage amplitude of current harmonics. V=220 V RMS of the fundamental term.

By means of the hammer test, it was possible to identify the most significant resonance frequencies of the machine in order 1, given the carrier wave frequency of 750 Hz and 50 Hz of the modulating wave. The vibration graph shows after the test that the resonance frequencies coincide with the calculations made with equations (1-6) with the construction parameters of the machine, which are as follows: Modulus of elasticity: $E_c$=200e9 Pa; density $\rho$=7700·9.8 N/kg; $D_c$=0.176 m; stator slots: $s_1$=36; tooth height: $h_t$=0.008 m; stator thickness: $h_c$=0.01 m; tooth width: $c_t$=0.0087 m; stator length: $L_i$=0.25; and stator diameter: $D_c$=0.186 m.

As can be clearly seen in Table 2 (calculated values) and Figure 11 (hammer test), there are resonance frequencies measured at 1600 Hz, which corresponds to the calculated 1571 Hz, as well as the calculated frequency of 2921 Hz (Table 2), which practically coincide with the experimental result. The calculated housing resonance frequencies exceed 10 kHz, outside the measurement range, and have therefore not been provided.

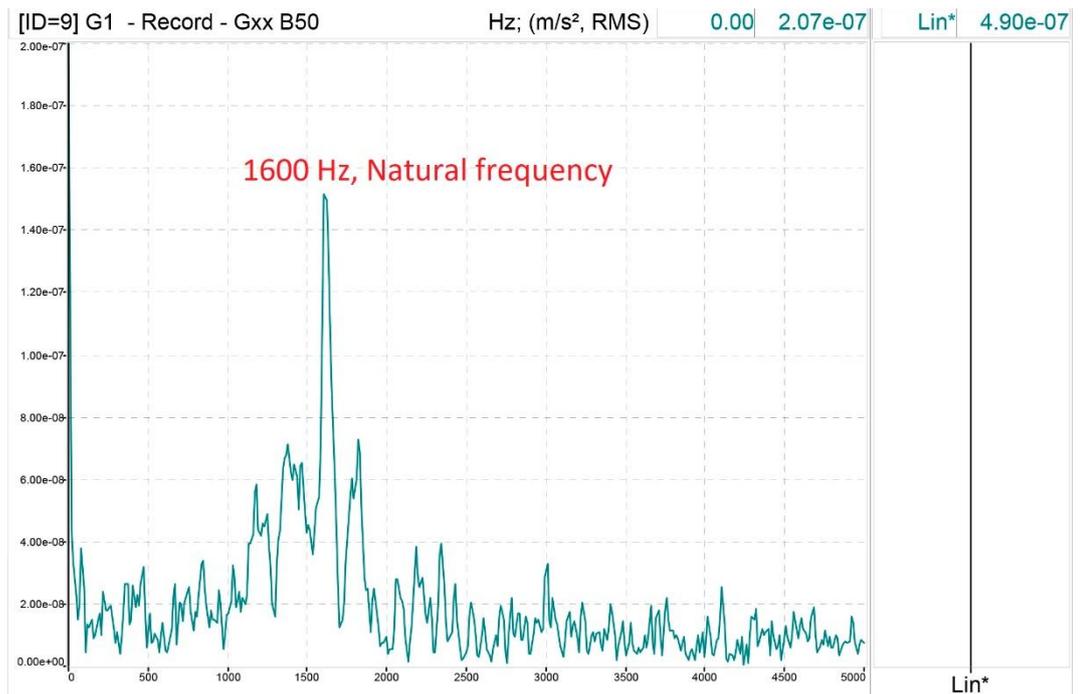

**Figure 11.** Natural Frequency Testing with an Impact Hammer acceleration m/s² versus frequency.

**Table 2.** Radial frequencies calculed for the induction motor.

| M | Frequency (Hz) | Frequency (Hz) | Frequency (Hz) |
|---|---|---|---|
|   | 0.45*e+02* | 0.229*e+03* | 0.547*e+03* |
| 1 | 0.995*e+03* | 0.1571e+04 | 0.2921*e+04* |
|   | 0.6533*e+04* | 0.8519 *e+04* | 0.4619*e+04* |
| 0 | 2.9206*e+03* | | |

The vibration results as a function of the frequency of the mains-fed motor and the multilevel inverter are shown below for SPWM-I, SPWM-II, SPWM-III and HIPWM-FMTCt techniques. This last technique will only be presented for K=0.55, for which the vibration level RMS is lowest. The RMS power supply value for all techniques was 220 V RMS. The average frequency of the carrier signals was 750 Hz.

Figure 12 shows the vibrations due to the spatial harmonics (tooth harmonics), which are the most significant as mentioned in Section 2 and its are clearly distinguishable with the sinusoidal power supply. The frequency of the spatial harmonic will be $\nu \pm 1$ as With the motor used, the harmonic frequencies of the vibrations produced by these tooth harmonics appear above values: 18±1 (17,19), and 36±1 (35,37) with p=2, $s_1$=36. These vibrations will appear in all the graphs, since they are a consequence of the constructive characteristics of the machine. For the 50 Hz power supply, vibrations are identified for 900 Hz and 1800 Hz.

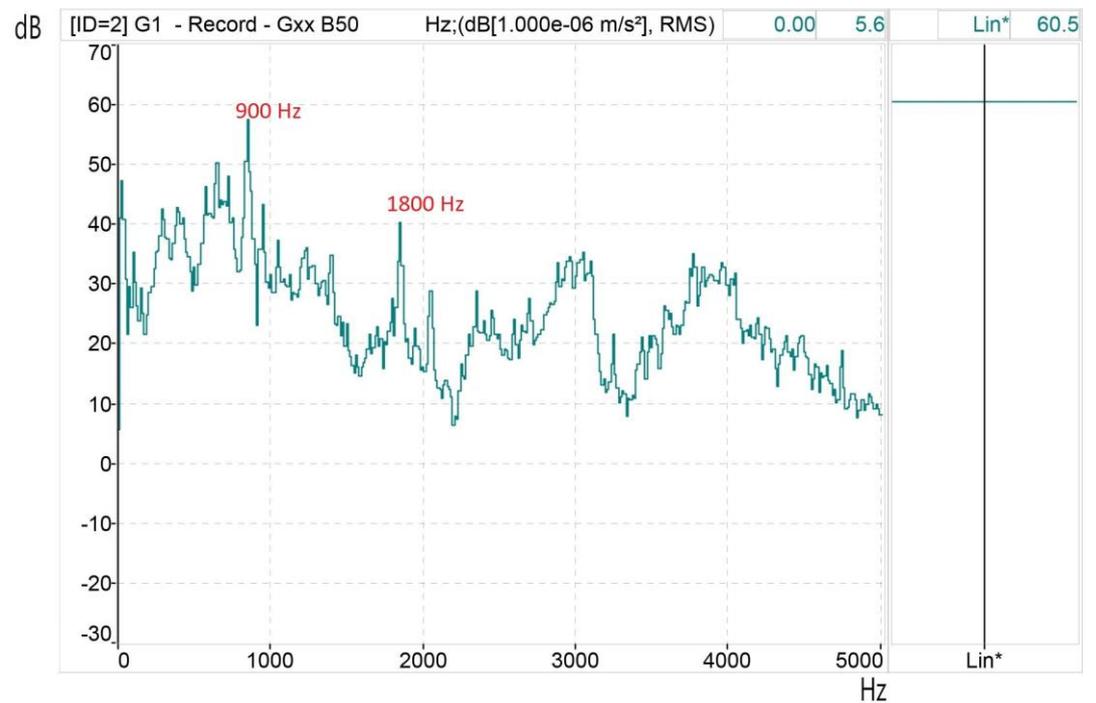

**Figure 12.** Vibration response -dB/Hz- of the electrical network.

Figure 13 shows the results of the SPWM-I strategy with M = 15. In addition to the vibrations caused by using sinusoidal power supply, time harmonics also can be identified. It can be observed electrical harmonics 19 (1·750+2·50 Hz), 37(2·750+7·50 Hz), and 31 (1550 Hz), that activates the natural resonance of 1500 Hz. These electrical harmonics generate the vibrational frequencies: 1000 Hz, 1500 Hz and 1800 Hz, respectively, if the most significant ones are indicated.

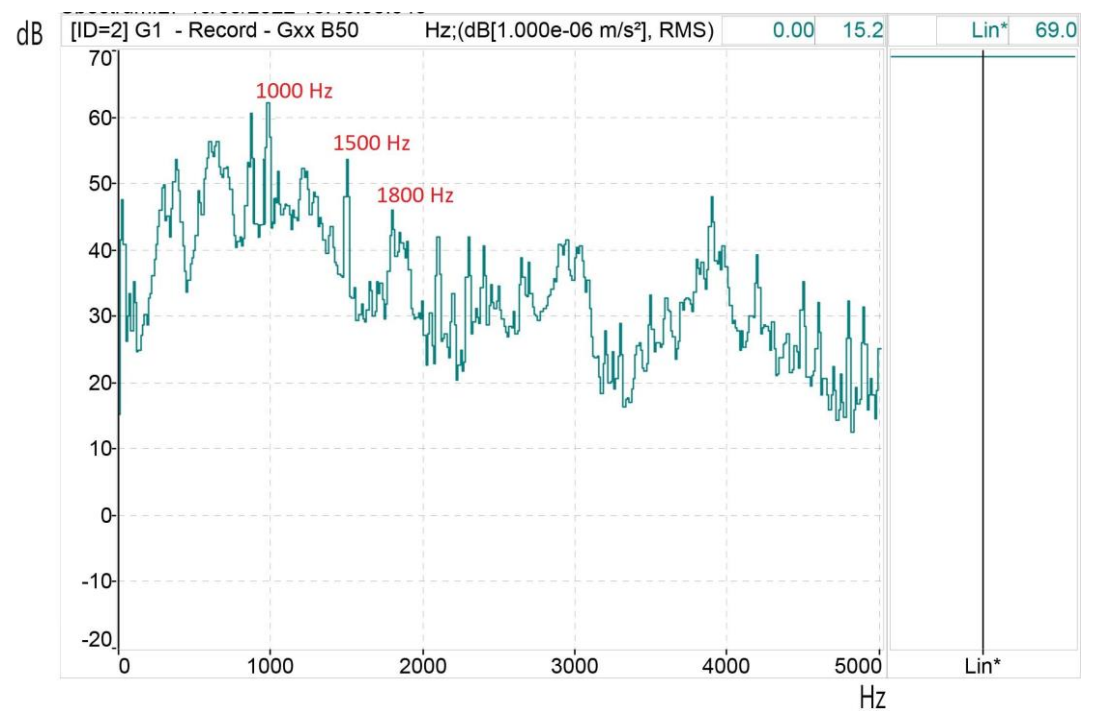

**Figure 13.** Vibration response -dB/Hz- for SPWM-I.

Figure 14 despites the results of the SPWM-II strategy with M = 15. In addition to the vibrations caused by space harmonics, time harmonics also appeared. It can be observed electrical harmonics 11, 17 and 23 and 29 that activates the natural resonance of 1500 Hz. These electrical harmonics generate the vibrational frequencies: 600 Hz, 900 (this harmonic coincides with one of harmonic of tooth), 1500 Hz and 1900 (second order tooth harmonic).

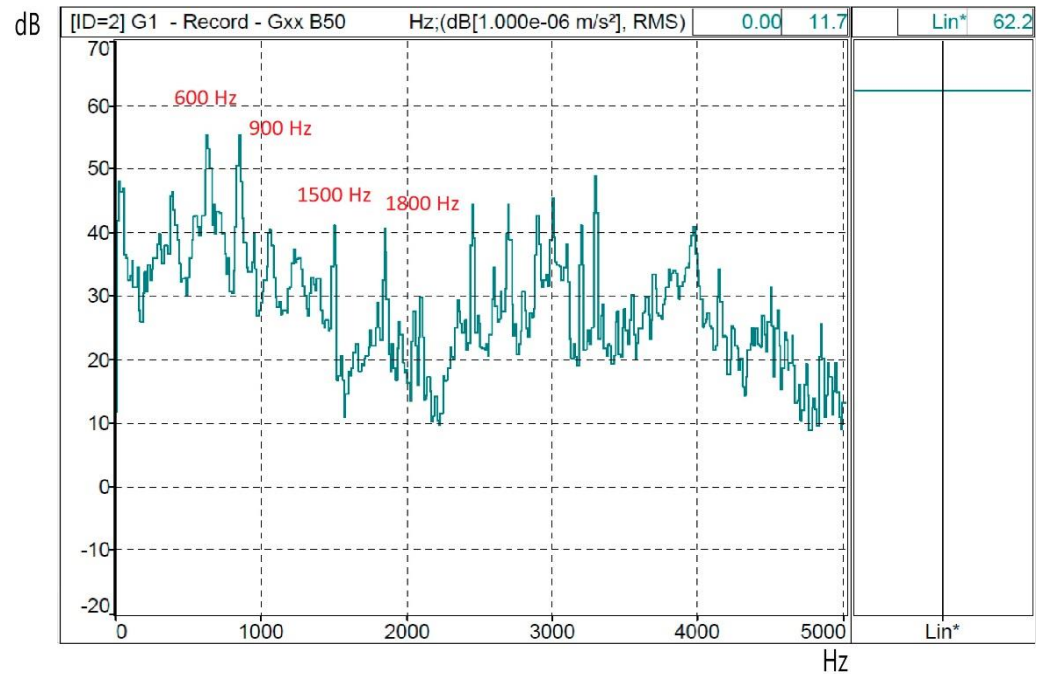

**Figure 14.** Vibration response -dB/Hz- for SPWM-II.

In Figure 15, we find in the results for the SPWM-III technique a vibration spectrum as a function of frequency very similar to the previous one. In addition to the vibrations caused by space harmonics, time harmonics also appeared. It can be observed electrical harmonics 11, 17 and 23 and 29 that activates the natural resonance of 1500 Hz. These electrical harmonics generate the vibrational frequencies: 600 Hz, 900 (this harmonic coincides with one of harmonic of tooth), 1500 Hz and 1900 (second tooth harmonic).

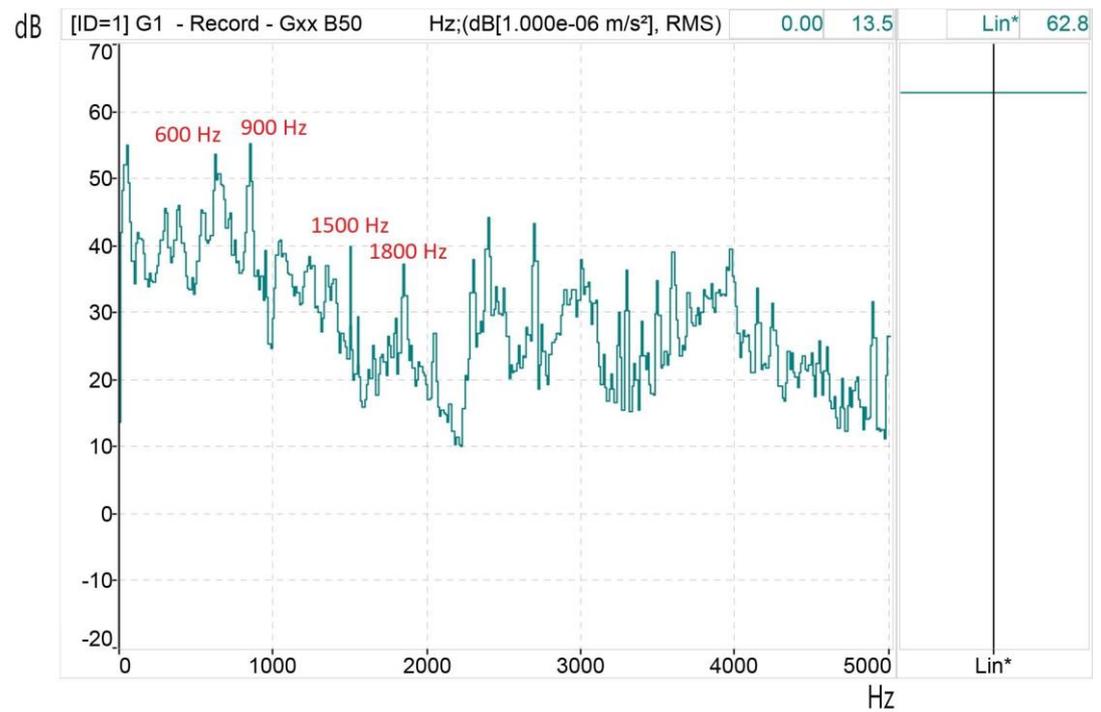

**Figure 15.** Vibration response -dB/Hz- for SPWM-III

Finally, Figure 16 depicts the results of the HIPWM-FMTCt proposed technique with K=0.55 and $\bar{M}$ =15. It can be seen that vibrations similar to the three previous ones appear but the vibration is lower and the natural resonance frequency of 1500 Hz has not been activated.

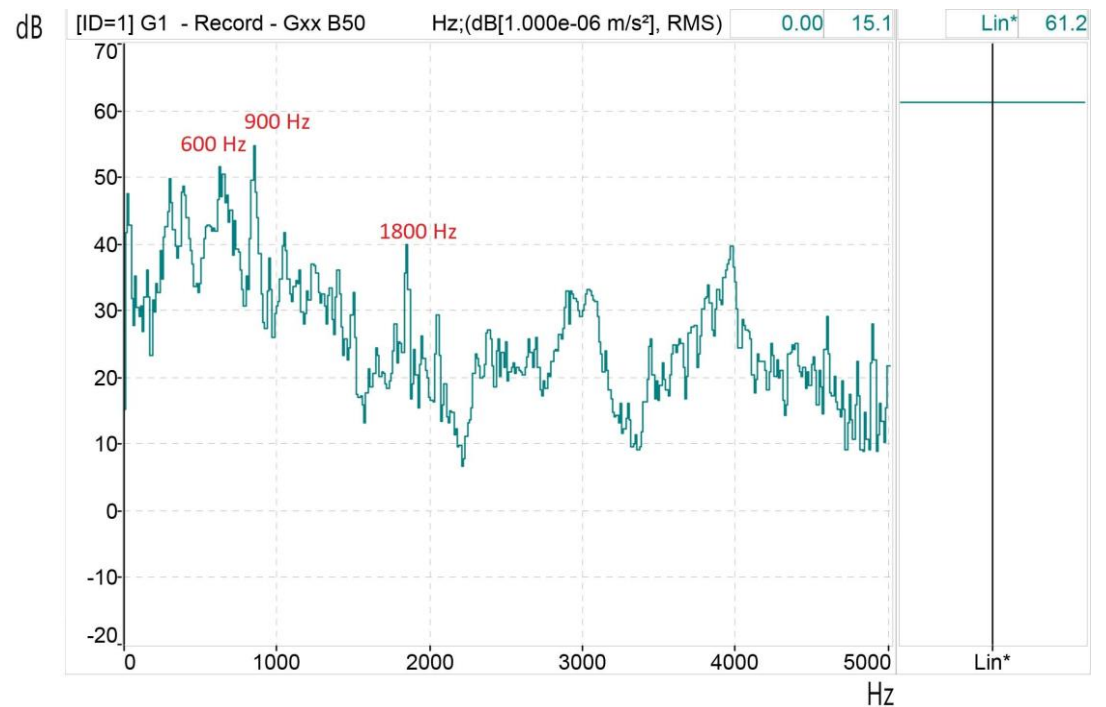

**Figure 16.** Vibration response -dB/Hz- for proposed technique with K=0.55

Table 3
Experimental Results for $\bar{M}$ =15 and M=15

| Type of PWM/Power Supply | Level of Truncation | Level of vibrations dB 220V RMS fundam. term | THD Voltage (%) 220V RMS fund. Term | $V_{RMS}$(V) DC-link 75V |
|---|---|---|---|---|
| Power supply | | 60.5 | - | - |
| SPWM-I | | 69.0 | 15.0 | 187 |
| SPWM-II | | 62.2 | 4.52 | 190 |
| SPWM-III | | 62.8 | 5.73 | 226 |
| HIPWM-FMTCt | K=0.3 | 63.1 | 8.58 | 220 |
| | K=0.4 | 62.9 | 7.05 | 223 |
| | K=0.45 | 62.8 | 4.01 | 226 |
| | K=0.5 | 62.6 | 4.25 | 231 |
| | K=0.55 | 61.2 | 4.51 | 229 |
| | K=0.6 | 62.3 | 5.03 | 228 |
| | K=0.65 | 62.7 | 5.24 | 230 |
| | K=0.7 | 63.0 | 5.52 | 230 |
| | K=0.75 | 63.2 | 6.87 | 231 |
| | K=0.8 | 63.8 | 7.39 | 231 |

## 5. Conclusions

A frequency modulation based PWM technique, synchronized with the modulating wave, was simulated and implemented in a cascaded multilevel inverter. Laboratory measurements were performed to evaluate both vibrational and electrical results. The advantage of the proposed HIPWM-FMTCt technique lies in its ability to modify the electrical characteristics of the inverter output waveform and, consequently, the vibration results in a machine, using a control parameter K with an equal number of switching pulses. Depending on the optimization objective, be it minimizing vibrations, reducing total harmonic distortion (THD) or increasing the root mean square (RMS) value of the output voltage, the most suitable K value can be determined. This information is presented in Table 3 of the document, which shows the effects of different values of K. For a K value of 0.55, vibrations are minimized. When K is set to 0.5, there is an improvement in THD, while higher K values result in increased RMS values at the inverter output. The main target of this study was to select K values that minimize the magnetic harmonic field (MMF) of the motor, thus reducing the vibrations caused by the induction motor. This purpose was achieved without significantly affecting other electrical parameters. Within the K range of 0.5 to 0.6, lower vibrations were observed, with the optimum value K=0.55, outperforming other modulation techniques described in the technical literature. The technique effectively reduces vibrations by avoiding mechanical resonances and spatial harmonics with a high winding factor. In addition, this technique is notable for having lower THD values than the techniques with which it was compared.

**Author Contributions:** A.R.G. conceptualization and methodology, F.P.H., J.H.L. and M.M.G. Simulation and Data collection, A.R.G., F.P.H., J.H.L. and M.M.G hardware implementation, A.R.G. writing—original draft preparation A.R.G., F.P.H., J.H.L. and M.M.G validation, writing—review and editing. F.H.P. funding acquisition. the authors have read and agree with the final version of the manuscript.

**Funding:** This research was funded by National Plan Project ENE-19744C0302, Spain.

**Conflicts of Interest:** The authors declare no conflict of interest.